\documentclass[a4paper,fleqn]{cas-dc}

\usepackage[authoryear]{natbib}
\usepackage{graphicx}

\usepackage{float} 
\usepackage{subfig}

\begin{document}
\let\WriteBookmarks\relax
\def\floatpagepagefraction{1}
\def\textpagefraction{.001}

\shorttitle{Exit Dynamics of a Square Cylinder} 

\shortauthors{I.Ashraf et~al.}

\title[]{Exit Dynamics of a Square Cylinder}
\tnotemark[]

\tnotetext[2]{The financial support of the Belgian Fund for Scientific Research under research project WOLFLOW (F.R.S.-FNRS, PDR T.0021.18) is gratefully acknowledged. Part of the experimental setup was financed by {\it Fonds Sp\'eciaux} from ULi\`ege. SD is F.R.S--FNRS Senior Research Associate.}


%
\author[1]{Intesaaf Ashraf}[orcid=0000-0002-1405-3803]

\cormark[1]

\fnmark[]

\ead{intesaaf.ashraf15@gmail.com}



\affiliation[1]{organization={CESAM--GRASP, Physics Department B5, University of Liege},
            addressline={}, 
            city={Liege},
            postcode={4000}, 
            state={},
            country={Belgium}}

\author[1]{Stephane Dorbolo}[]

\cortext[1]{Corresponding author}



\begin{abstract}
In this paper, we experimentally investigate the exit dynamics of a square cylinder, that is initially fully immersed in a water tank and that crosses the interface perpendicularly to its symmetry axis.  The cylinder moves upwards at a constant velocity in the vertical direction till the cylinder exits out of the water into the air.  The experiments were performed at different traveling speeds.  The images of the cylinder crossing the interface were taken using a high-speed camera.  The images were used to track the interface deformation when the cylinder approaches and crosses the interface. On top of these measurements, the force required to move the cylinder was simultaneously measured in order to estimate the drag force during the travel in the tank, the force of entrainment, and the force of crossing over the interface. Particle image velocimetry was performed to visualize the flow. Correlations between the different measurements are inspected.

\end{abstract}



\maketitle
\section{Introduction}
\label{intro}
The interaction of a rigid body with the free water surface, when it enters or exits a fluid, is important in numerous domains, including ocean and coastal engineering, aviation industries, naval architecture, and dip-coating research. The situation's complexity makes physical interpretation and mathematical modeling a difficult task that involves fluid-structure interaction, interfacial physics, and fluid mechanics.

There have been many research studies that address the physics of the entry of an object in a fluid   \citep{challa2014,challa2010,mohtat2015,yang2012} and exit of an object from fluid \citep{Havelock1936,truscott2016,wu2017experimental,haohao2019numerical}. Compared to the extensive investigation on the water entry problem, the water exit has been far less studied than the water entry. To expand our understanding of the water exit phenomena, validate numerical simulations, and contribute to the development of new analytical models, new experimental research is required. \citet{Havelock1936} provided an analytical solution of the vertical motion of a cylinder placed in a uniform stream. 

\citet{greenhow1983nonlinear} took an experimental approach to the exit dynamics of a neutrally buoyant cylinder originally positioned on the bottom of a water tank and driven upward while exerting a constant force. According to their findings, the upward movement induced an elevation of the water-free surface, resulting in a bump-like formation. This bumpy pattern eventually breaks down in a chaotic way, which is referred to as ``waterfall breaking''. \citet{telste1987} investigated the exit of a circular cylinder through a computational method using potential flow theory. It was found that when approaching a free surface at low- speed, the fluid behaves as if the cylinder is approaching a wall. At high speeds, however, the cylinder acts as if it was traveling through an infinite fluid. The surface wave motion was observed at intermediate speeds.

Among the other research made on the exit of an object from a fluid, \citet{greenhow1997water} employed a two-dimensional numerical simulation to study the forced motion of totally and partially submerged horizontal circular cylinders. \citet{liju2001} addressed the upwards and downwards movements of various axisymmetric cylindrical bodies at different Reynolds numbers and studied the surface surging effect. \citet{moshari2014} used the VOF (Volume of Fluid Method) model, to simulate in 2D and 3D the water exit of a circular cylinder. They reported the formation of waves, wave motion in the horizontal direction, and air entrapment in the oblique exit. In the literature, we can also find  \citet{kleefsman2004improved} and \citet{nair2018water}  whose work implemented an improved VOF method to study the exit of a cylinder from water. \citet{chu2010} studied the water exit of a cylinder, in which the movement in the longitudinal direction of the cylinder was performed. In this work, they observed cavities at both ends of the cylinder. The slapping of the water was observed due to the collapse of those cavities. They performed experiments for both the accelerating and decelerating motion of the cylinder.  

\citet{haohao2019numerical} simulated the exit dynamics of a sphere moving at a constant velocity using LBM (Lattice Boltzmann techniques). They reported that the elevation of the water surface is strongly dependent upon the Froude number below 4.12. However, the Froude number ($Fr=\frac{U^2}{ga}$, whereas $U$ is the vertical speed of the object, $a$ is the characteristic length of the object and $g$ is the acceleration due to gravity) ranging between 4.12 and 8.24, slightly affects the free surface elevation height. The waterfall breaking becomes more intense and affected as the Froude number increases. The Reynolds number ($Re=\frac{Ua}{\nu}$, whereas $U$ is the speed of the object, $a$ is the characteristic length of the object and $\nu$ is the kinematic viscosity of the fluid) dominates the flow when the sphere moves beneath the water's surface. Through the simulation of a water exit of a fully submerged spheroid,  \citet{ni2015} reported that the movement at which the free surface breaks up from the body can be delayed by making the object blunter.   
\citet{truscott2016} conducted experiments on the behavior of a buoyant sphere that rises to the surface under the influence of buoyancy and eventually pops out of the water. They came to the conclusion that Reynolds number has a significant impact on the creation and shedding of vortices. The sphere's trajectory can be either a straight line or an oscillating path, depending on the release depth. 

\citet{wu2017experimental} investigated the free surface deformation and its dependency on the velocity or Froude number. It was done for both fixed and free spheres ascending towards the water surface. \citet{wu2017experimental} looked at free surface deformation and its correlation with velocity and Froude number. Experiments were performed for both fixed and free spheres as they ascended to the water's free surface. They discovered that the height of water elevation increases with body velocity (or equivalently Froude number) for a fully immersed sphere. In the case of a partially submerged sphere, the water detached from the sphere's surface in the form of a water column at the end of the exit stage.

When we look at all of the studies in the literature, we can see that they are mostly concerned with the waterfall breaking, pop-up heights and its trajectory. Less is knwon about forces during the crossing and  drainage, both of which are important in engineering applications. The shape is certainly a crucial parameter that influences on the exit dynamics of an object. Almost all the objects in these investigations were spherical or cylindrical. 

In the present study, we aim to address this problem by investigating the exit dynamics of a square cylinder pulled out of the water at a constant speed. As the square cylinder was pulled out of the water at a constant speed, the interface deformation was measured in order to inspect the evolution of the thickness of the water layer on top of the cylinder.  The force measurements were synchronized with the images taken by two high-speed cameras. Two non-dimensionalized numbers are used in the experimentation. i.e. Froude number and Reynolds number. The Froude number is defined as $Fr=\frac{U^2}{ga}$, where $U$ is the vertical speed of the cylinder and $a$ is the side of the cylinder. And, the Reynolds number is defined as $Re=\frac{Ua}{\nu}$ where $\nu$ is the kinematic viscosity of the water. We focus on the crossing of the interface. Indeed, several phenomena have to be taken into account to model the quantity of liquid entrained out of the bath such as the speed and the shape of the object on one side and the surface tension and the viscosity of the fluid on the other side. Here we focus on the shape of the object.

\section{Experimental Setup}

The experimental setup consists of a lifting system that pulls the test object, a glass water tank (length = 78.5 cm x width = 27.5 cm x depth = 72.5 cm), a force sensor, and a test object.

\begin{figure*}[H]
	\centering
	\includegraphics[scale=0.65]{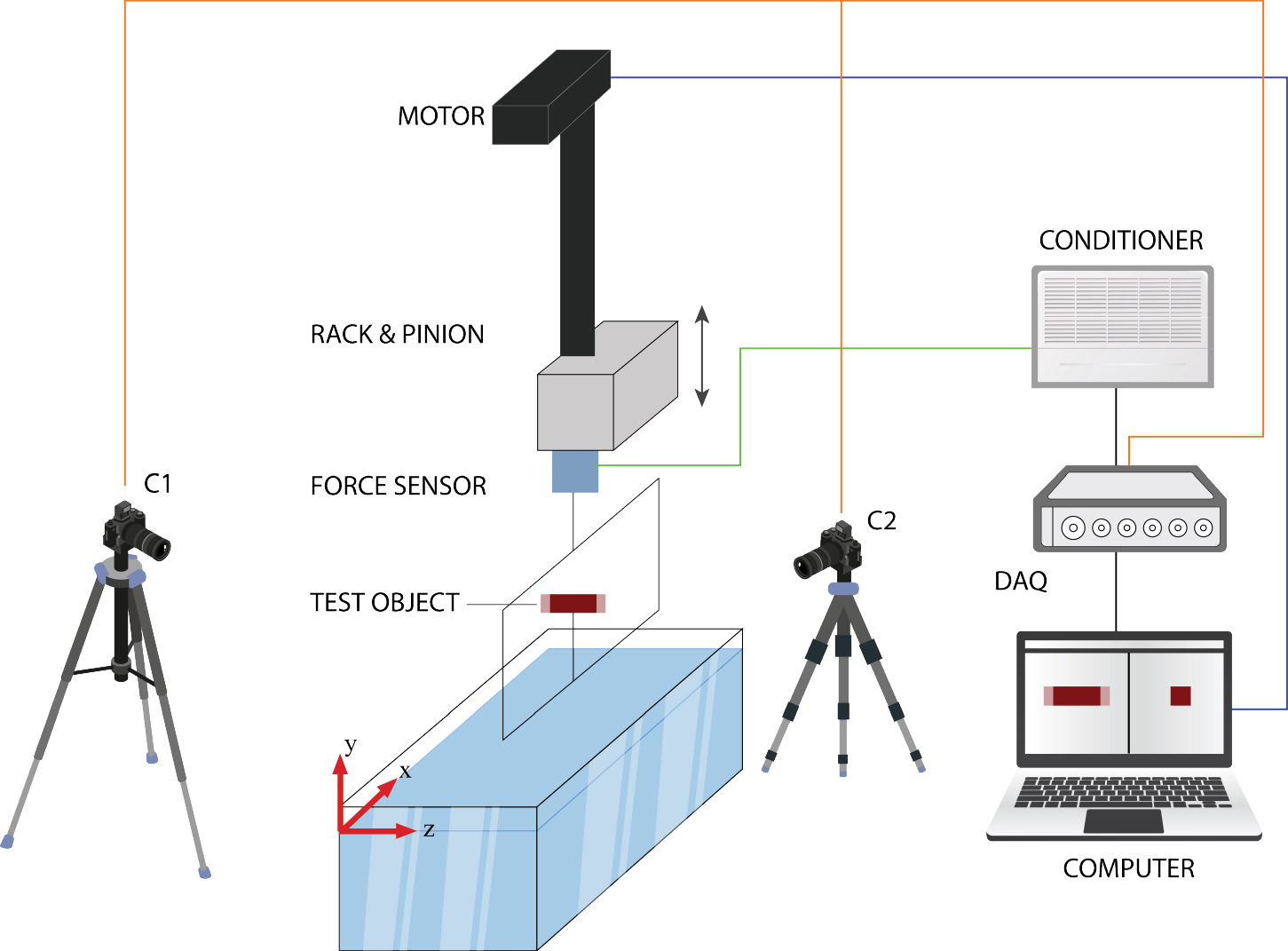}
	\caption{The schematic of the experimental set-up. The square cylinder was screwed into a carbon fiber frame. The frame was attached to the force whereas the force sensor was attached to the rack and pinion mechanism. C1 and C2 were two high-speed cameras used to record the front and side views of the motion.}
	\label{setup}
\end{figure*}

The lifting system uses a rack and pinion Fig. \ref{setup}, the moving part of the rack was attached to a frame made of carbon fiber tubes, which held the cylinder horizontally, parallel to the water tank's side and front walls.

A right square cylinder having square bases of side $a= 40mm$ and rectangular faces of height $h=220$ mm was 3D printed and used for the experiments. 


The cylinder was placed horizontally so that its square bases were always perpendicular to the water surface. It was attached to the mobile frame with a rod bolted to one point of its vertical symmetry plane of the cylinder.

The coordinate system ($xyz$) is shown in Fig. \ref{setup}. The $x$-axis is oriented along the length of the tank, and the $z$-axis is aligned along its width. The origin is set at the free water surface, which is the horizontal reference plane ($xz$). The pulling axis, aligned with the vertical direction, is defined as the $y$-axis. The coordinates are normalized by dividing them by $a$, the side of the cylinder’s square bases.

The experiments were performed at a constant terminal velocity ranging from 0.1 to 1 m/s. The acceleration of the motion was set to 4 m/s$^2$. This means that the test object must travel a certain distance from rest before reaching its constant velocity. For example, to reach a constant speed of 1 m/s, the test object must travel 120 mm. The same applies to the deceleration phase.

Two high-speed M-310 phantom cameras were used to acquire images at a rate of 2000 Hz. Their field view was perpendicular to allow complete visualization of the crossing. The cylinder position was tracked using Matlab code developed in-house. To eliminate parallax and allow tracking of the interface deformation, the air-water interface was placed in the middle of the image.

A strain gauge SCAIME K25 (20 N) connected between the frame and the rack measures the forces exerted on the cylinder during its vertical movement at the same time as the imaging. A datalogger (Picolog) was used to record the data at a frequency of 1 kHz. The Picolog also recorded the camera trigger signal, allowing the force measurements to be synchronized with the images. This synchronization was used to determine the precise moment at which the object crosses the interface to calibrate the timeline of the force measurements.

The Particle Image Velocimetry (PIV) technique was adopted for 2D planar flow quantification. An illumination laser with a wavelength of 532 nm and peak power of 4W was used for exploration together with 4W for exploration together with tracer particles having a size equal to $\simeq$ 20$\mu$m.The PIV images were post-processed using the open-source software PIVLab \citep{thielicke2014pivlab}.

\section{Result and Discussion}
The cylinder was first placed in the tank at a depth $d = 8.15a$ below the free water surface, and the system was left to rest for 20 minutes. After ensuring that all surface waves were damped, the cylinder was lifted upward at a constant speed. As the cylinder moves upward, it deforms the free surface of the water. The maximum deformation of the surface is achieved when the cylinder reaches the water surface.  



\begin{figure*}
\centering
\subfloat[][]{\includegraphics[width=0.65\textwidth]{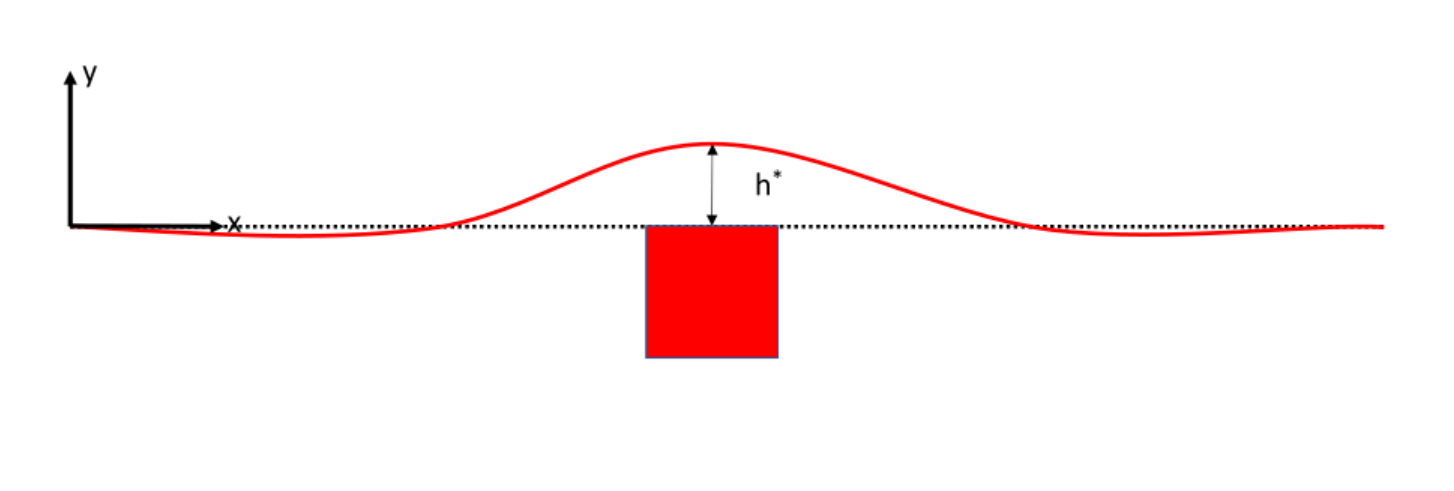}}\\
\subfloat[][]{\includegraphics[width=0.65\textwidth]{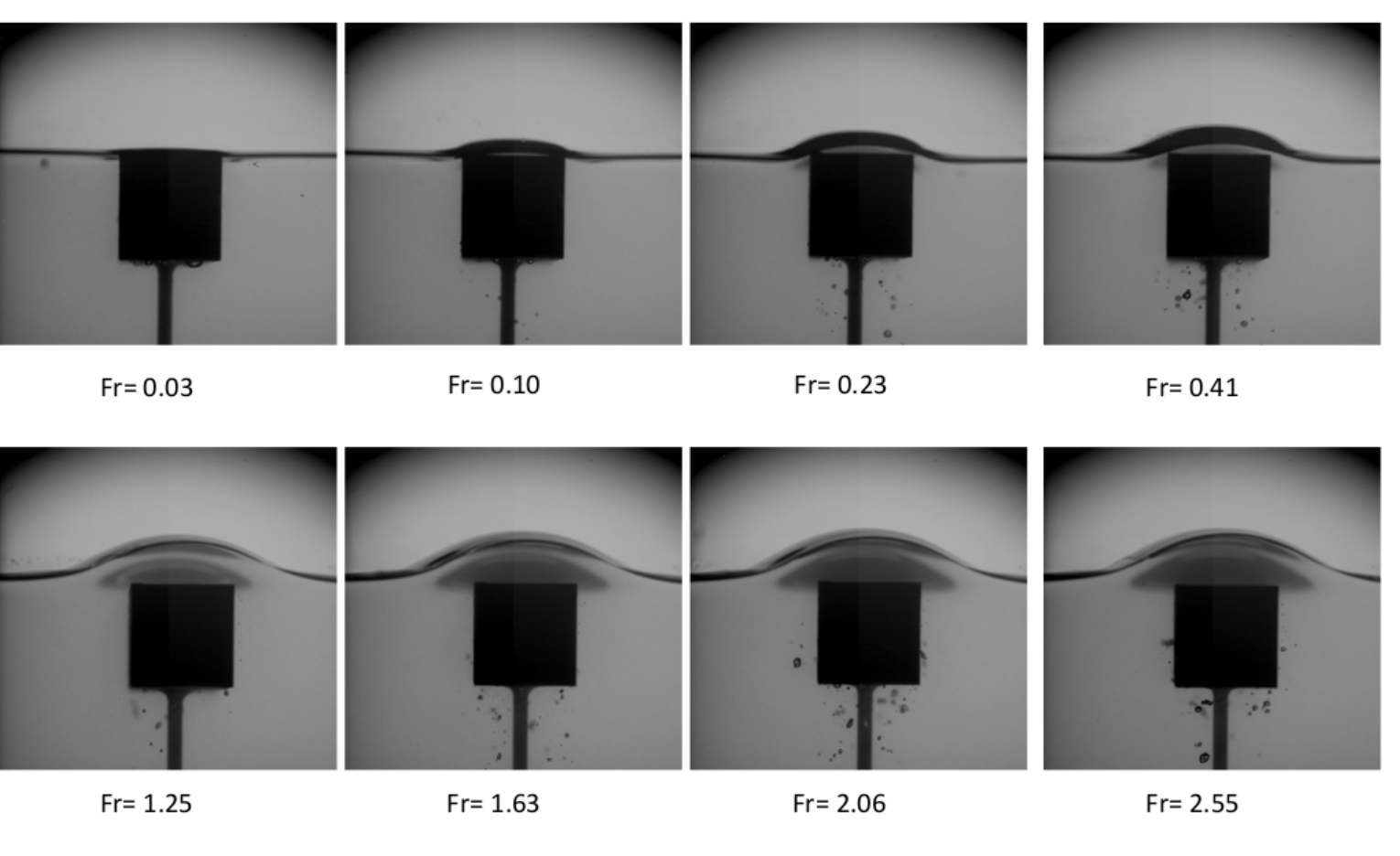}}\\
\subfloat[][]{
  \includegraphics[width=0.65\textwidth]{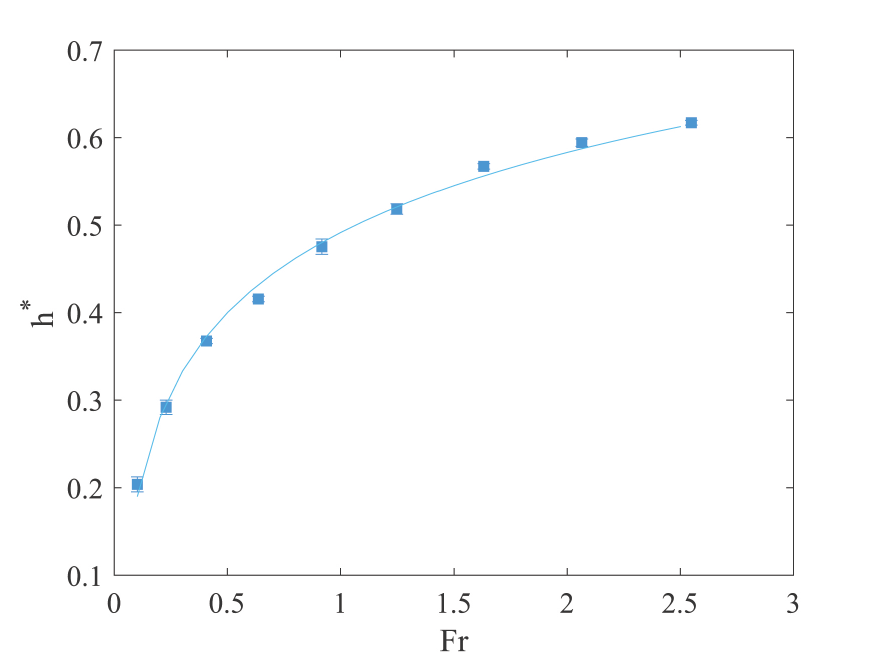}}
\caption{(a) Scheme illustrating the maximal deformation. $h^*$. It is non-dimensionalized by the side of the square, $a$. (b) Pictures showing the free surface elevation at maximal deformation, when the cylinder is at a position:  $y_c = -1/2$ for eight different Froude Numbers. (c) Evolution of  $h^*$ as a function of Froude number, Fr. The symbol is average data over 7 experiments and error bars are standard deviation. The curve fitted is a  fit  ($h^*=\alpha ln(Fr)+h_o$), where  $\alpha = 0.1321$ and $h_o = 0.5$. }
\label{fig:2}       
\end{figure*}

The maximum deformation $h^*$ (non-dimensionalized by $a$) can be measured to characterize the water entrainment. It is defined as the initial film thickness or the elevation of the liquid level at the movement when the cylinder starts to cross the interface, namely when $y=-1/2$. It is determined by measuring the height of the water-air interface from the reference plane, as shown in Fig.\ref{fig:2}a.
The maximum deformation $h^*$ is related to the entrained volume of water above the free surface since it represents the height of the liquid above the cylinder when the top of the cylinder reaches the initial level of the free surface.

In the first set of experiments, we investigated the effect of the cylinder's velocity or Froude number on $h^*$. Fig. \ref{fig:2}b shows images of the maximum deformation at this movement when the position of the cylinder is $y_c$ = $- 1/2$, for different Froude numbers. In addition, Figure 2(c) shows the values of $h^*$ calculated from eight experiments for eight different Froude numbers.

Fig. \ref{fig:2}b and Fig. \ref{fig:2}c show that the entrainment of water increases as the Froude number increases. At low velocities, $h^*$ is very low. Then, it increases with the Froude number and reaches a value of about 0.6 at Froude number Fr = 2.54, i.e., 0.6 times the side of the cylinder.

\begin{figure*}
	\centering
	\captionsetup{justification=centering}
	\subfloat[][]{\includegraphics[width=0.65\textwidth]{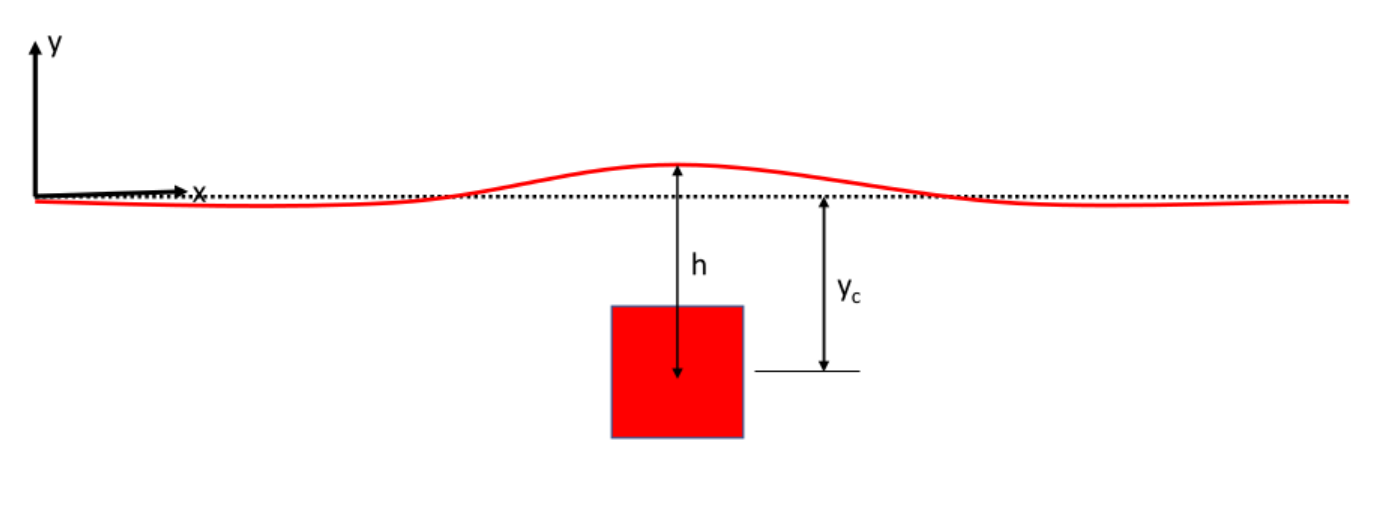}}\\
	\subfloat[][]{\includegraphics[width=0.65\textwidth]{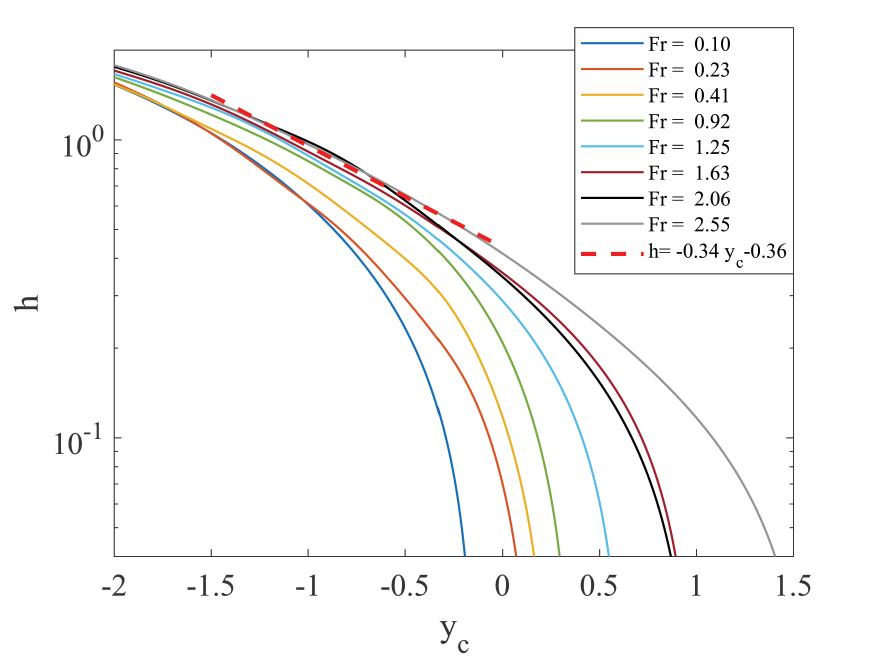}}
	\caption{(a) Definition of $h$ and $y_c$.  $h$ and $y_c$ is non-dimensionalized by the side of the square, $a$. (b) $h$ versus $y_c$ as a  function of Froude number, Fr.}
	\label{fig:3}       
\end{figure*}

The Fig. \ref{fig:3}b shows the evolution of surface elevation $h$ defined as the distance from the center of the square cylinder to the top of the free water surface (Fig. \ref{fig:3}a) as a function of $y_c$ the position of the square cylinder. All values are non-dimensionalized by $a$. The surface elevation $h$ decreases exponentially with $y_c$ for small lengths at low Froude numbers. However, at a high Froude number, the exponential decay increases. It basically means that the drainage is exponential for a certain distance of travel of the square cylinder. The curve has a slope of -0.34 for the Froude number greater than 0.92.  

Further, the wake behind the square cylinder, when the top of the cylinder reaches the interface, is investigated by using PIV. The wake consists of the vortex structures attached to the side of the cylinder. No vortex shedding is observed in the case of a square cylinder. Also, the flow separation begins at the leading edge.  


\begin{figure*}
	\centering
	\captionsetup{justification=centering}
\subfloat[][ Fr= 0.10, Re = 7724]{\includegraphics[width=0.25\textwidth]{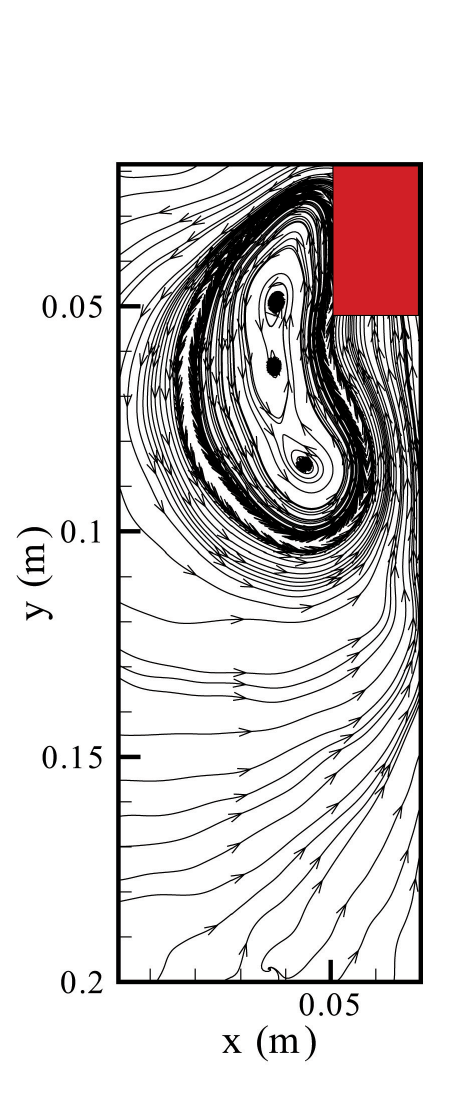}}
\subfloat[][ Fr= 0.92, Re = 23173]{\includegraphics[width=0.245\textwidth]{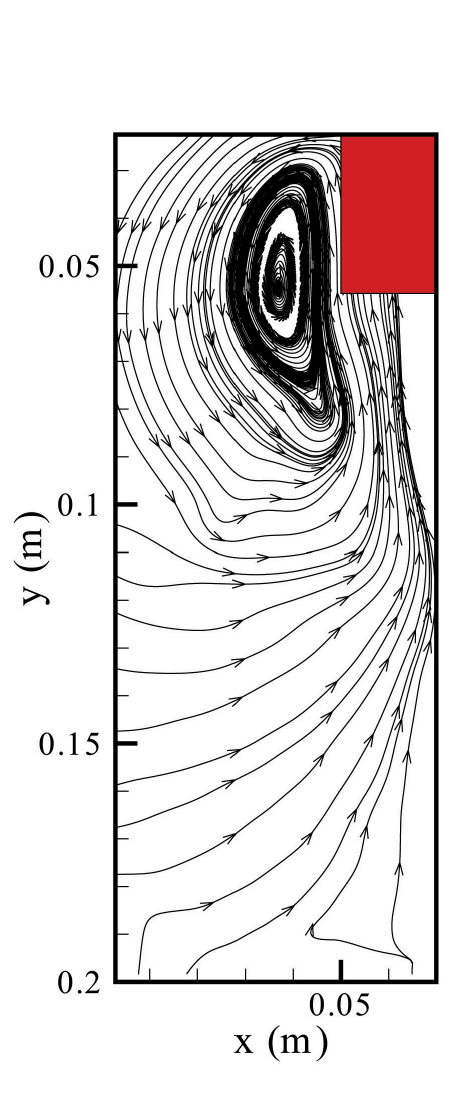}}
\subfloat[][  Fr= 2.55, Re = 38621]{\includegraphics[width=0.25\textwidth]{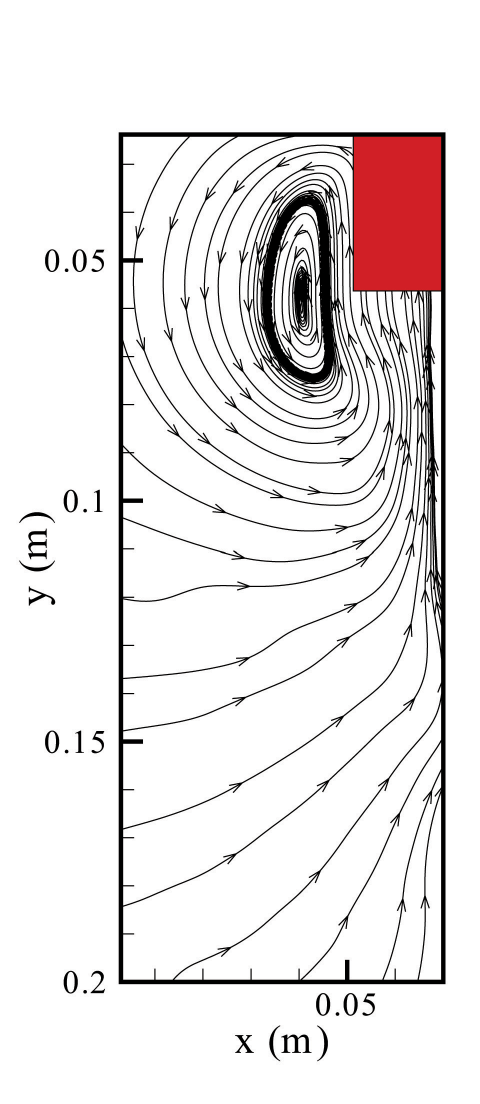}} \\
	\caption{Wake behind square cylinder as the cylinder rises to the water surface as a function of Froude number, Fr and Reynolds number, Re}
	\label{fig:6}       
\end{figure*}

\begin{figure*}
\centering
\includegraphics[scale=0.52]{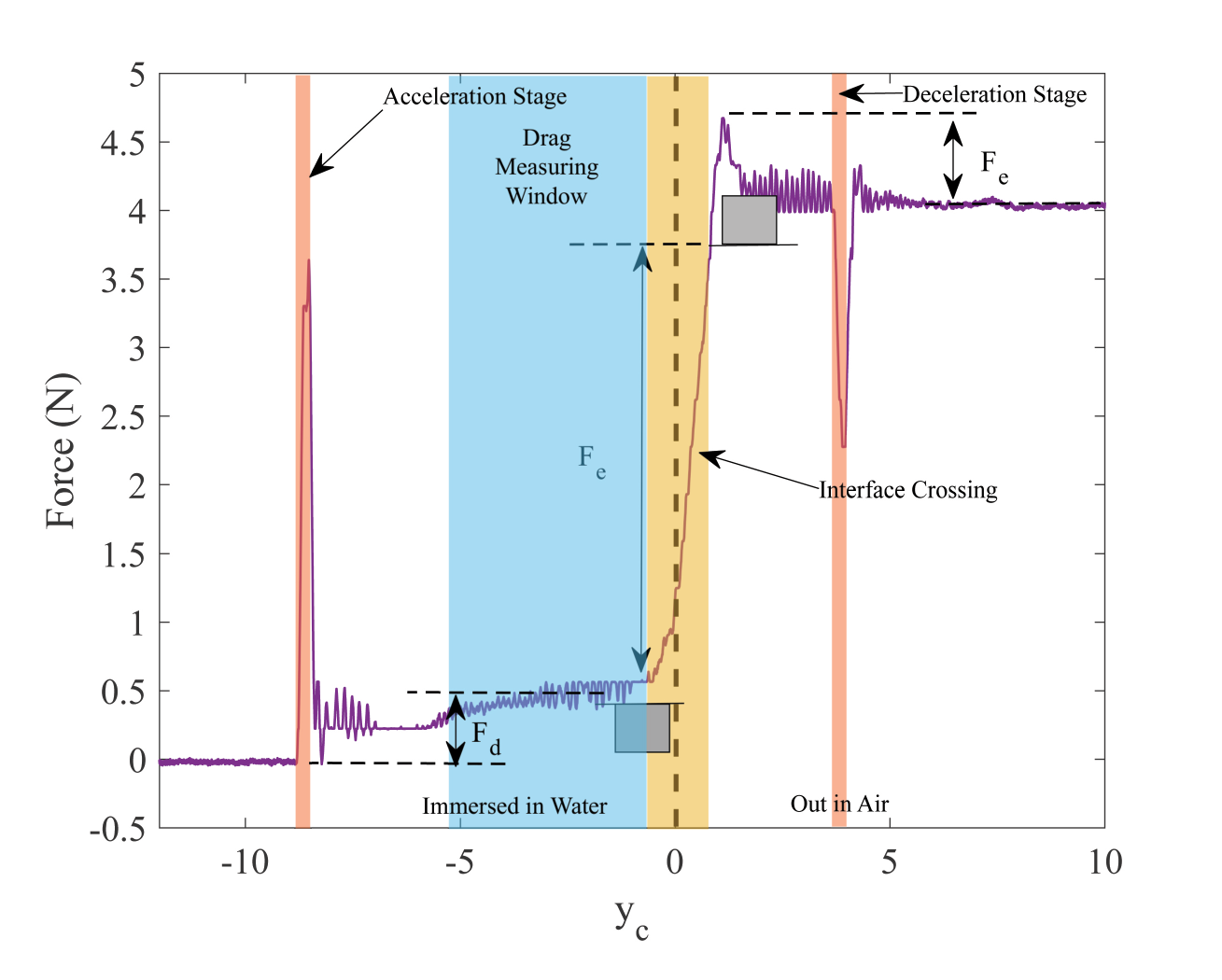}
\caption{Definitions stages of the cylinder in terms of force measurement  Fr=0.10. $y_c$ is non-dimensionalized by the side of the square, $a$.} 
\label{fig7}
\end{figure*}

\begin{figure*}
\centering
\includegraphics[scale=0.65]{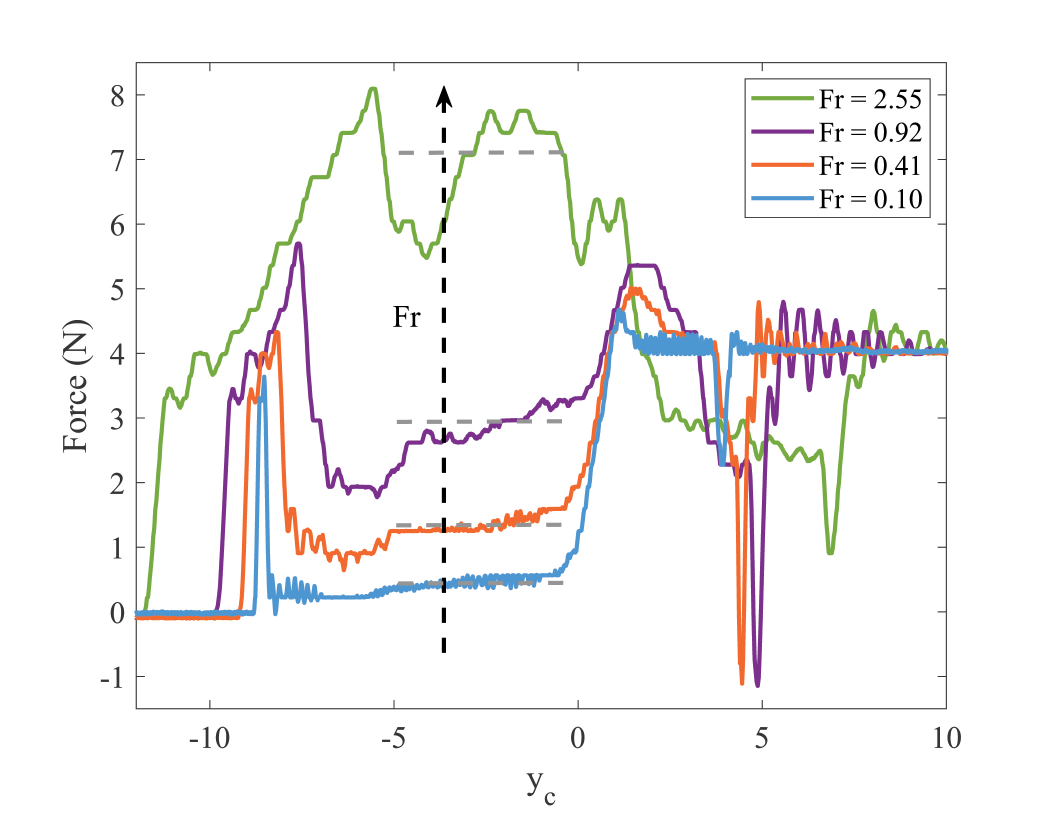}
\caption{Force experienced by the cylinder as a function of $y_c$ at different Froude numbers. $y_c$ is non-dimensionalized by the side of the square, $a$.} 
\label{fig8}
\end{figure*}

\begin{figure*}
	\centering
	\captionsetup{justification=centering}
	\includegraphics[scale=0.65]{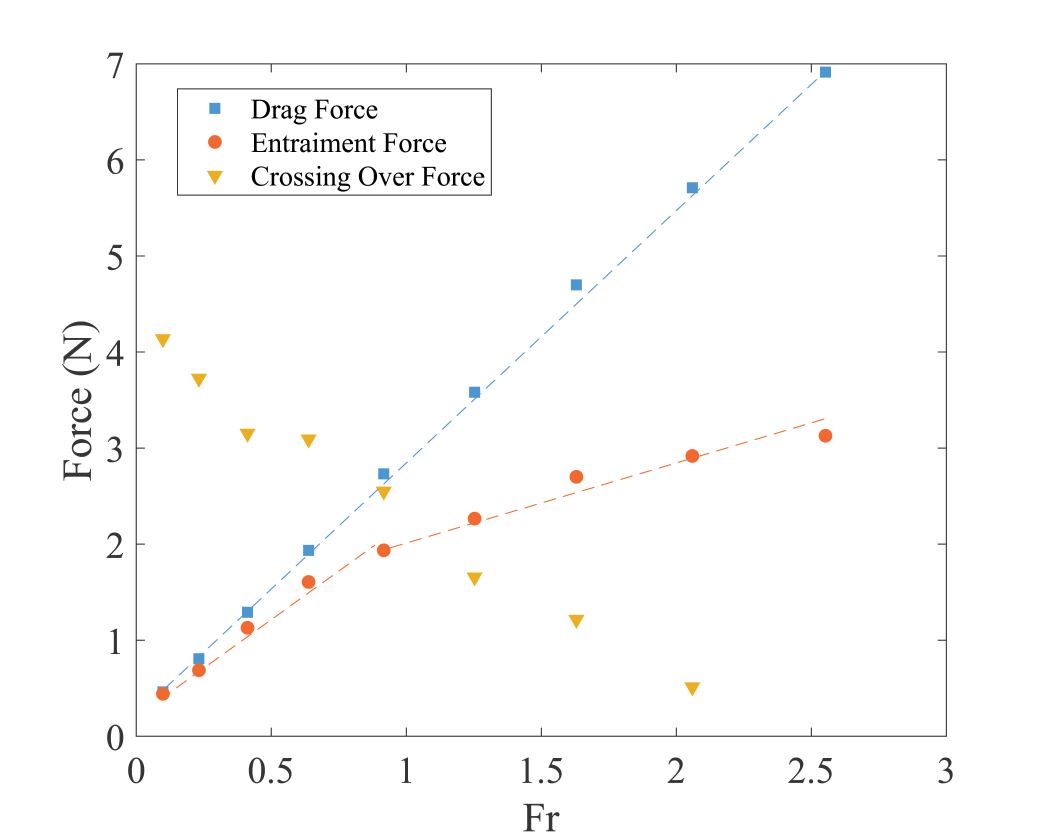}
	\caption{Drag Force (square), Entrainment Force (bullets) and Crossing over Force (triangle) as a function of Froude number.}
	\label{fig:9}       
\end{figure*}

The forces acting on the square cylinder during its upward movement and interaction with the free water surface were studied using a stain gauge sensor. As it has been carefully checked that the speed is constant during the cylinder motion, we can find the position of the cylinder by multiplying the time by the speed.  According to this procedure, what happens before the acceleration period and after the deceleration period can not be properly represented. 

To understand the registered force signal and examine the distinct phases of the cylinder's motion, the experiment is first conducted at low speed, for Fr=0.10.  Fig. \ref{fig7} shows the registered force variation as a function of depth. The cylinder motion is categorized into 5 zones or stages, as shown in the graph: 

(a)  The acceleration phase, highlighted in orange. The cylinder moves from rest to acquire the constant velocity priory set for each experiment. At the end of this stage, the square cylinder starts moving upward at a constant speed after traveling a short distance relative to the initial depth.
(b) The constant drag phase, highlighted in blue. The cylinder moves vertically in water at a constant speed. The cylinder experiences the drag force in this stage. The drag force $F_d$ acting on the square cylinder during this upward movement is measured during this phase. The net drag force is measured by taking the average force acting on the cylinder during this stage.  
(c) The crossing phase, highlighted in yellow. The cylinder speed is still constant at this stage. The cylinder starts crossing the interface when its top is in first contact with the free water surface, and the crossing ends when the bottom of the cylinder leaves the water surface at $y_c=-0.5a$. 
The crossing-over force, $F_c$ is measured when the square cylinder comes out of the water and its bottom is leaving the water surface. The crossing-over force is measured by taking the difference of force at  $y_c$ equals $-0.5a$ and $0.5a$ respectively.  
(d) The cylinder is completely out of the water and moving in the air at a constant speed. The entrainment force $F_e$ is calculated during this stage. It is equal to the force acting on the cylinder when it is completely out of the water minus the weight of the cylinder when it stops moving.  The cylinder comes to rest after a brief deceleration period.

The evolution of force acting on the cylinder as a function of $y_c$ at different Froude numbers is shown in Fig. \ref{fig8}. The gray horizontal line shows the region in which the drag force, $F_d$ is measured. It should be noted that at low Froude numbers, the force sensor readings  are  almost as flat as the horizontal line. Whereas at a high Froude number, it fluctuates between this horizontal line.  The variation of the net drag force $F_d$, the crossing over force $F_c$, and the entrainment force $F_e$ as a function of the Froude number are presented in Fig. \ref{fig:9}.
The drag force $F_d$ is found to increase linearly with the Froude number (the slope is 2.65). This result is expected as the drag force is proportional to the speed of the cylinder in a laminar flow regime, which is the case here.
The net entrained force $F_e$ is minimum at a low Froude number and increases with the increase of the Froude number. We can notice from the plot of $F_e$ versus Froude number presents two linear regions. In the first region, the slope is equal to 2.195 for $Fr < 0.91$ and it is equal to 0.7357 for $Fr > 0.91$. As the cylinder exits at a higher Froude number, it carries more momentum. This increased momentum creates a stronger suction effect, which, in turn, draws more surrounding fluid into the wake of the cylinder. Also, a higher Froude number indicates a regime, where inertial forces are more dominant than gravity. As a result, at a higher Froude number more liquids are sucked in and carried away by the cylinder. 

The behavior of the crossing force $F_c$ is also illustrated in Fig. \ref{fig:9}. Unlike $F_d$ and $F_e$, $F_c$ decreases with an increase in the Froude number. The crossing force is the net force acting on the cylinder out of the buoyancy, drag, weight and the component of surface tension in the vertical direction. All the forces except the drag force are independent of the Froude number. The drag force increases with an increase in the Froude number. Therefore, the net crossing-over decreases with an increase in the Froude number.

\section{Conclusion}
As soon as the cylinder moves upward at a constant speed, the surface starts getting elevated. Also, the maximum elevation is achieved, when the top of the cylinder touches the water surface. However, this maximum surface elevation is a function of the cylinder Froude number. The maximum surface elevation $h^*$ has logarithmic relation with the Froude number. 

Another stage of our experimentation process, the PIV experiments shows that as soon as the cylinder moves upward it formed clockwise and anti-clockwise vortices attached to its surface. However, no vortex shedding such as Von Karman Street is observed in the wake. The flow separation starts from the leading edge of the cylinder.

We finally ended our experimental research by carrying out the synchronized force measurement and upward cylinder movement of the cylinder. We characterized different force regimes in the exit dynamics of a square cylinder. From the force measurement, the net drag force, entrained force, and cross-over force are estimated. The net drag force and entrained force increase with the increase in cylinder upward velocity. However, net cross-over force is found to decrease with an increase in velocity. 

The present study only investigated the exit dynamics of a square cylinder. A detailed study of different shapes could be an interesting avenue for future research.
\bibliographystyle{model1-num-names}

\bibliography{cas-refs}

\vskip3pt


\end{document}